\documentclass[aps,pre,onecolumn,superscriptaddress,notitlepage,longbibliography]{revtex4-1}
\usepackage[dvipdfmx]{graphicx}% Include figure files
\usepackage{dcolumn}% Align table columns on decimal point
\usepackage{bm}% bold math
\usepackage{amsmath}% added by Yamauchi
\usepackage{accents}
\usepackage{amssymb}
\usepackage{feynmp}
\usepackage{url}
\usepackage{setspace}
\usepackage{color}
\usepackage{pifont}
\usepackage{hyperref}

\newcommand{\revision}[1]{{{#1}}}

\newcommand{\beq}{\begin{equation}}
\newcommand{\eeq}{\end{equation}}
\newcommand{\beqa}{\begin{eqnarray}}
\newcommand{\eeqa}{\end{eqnarray}}
\newcommand{\Tr}{\mathrm{Tr}}

\newcommand{\ketbra}[2]{\left.\left| #1 \right\rangle\right.\hspace{-1.5mm}\left.\left\langle #2 \right.\hspace{-0.5mm}\right|}
\newcommand{\braketav}[3]{ \left\langle #1 \left| #2 \right| #3 \right \rangle}

\newcommand{\calL}{\mathcal{L}}
\newcommand{\calD}{\mathcal{D}}
\newcommand{\calH}{\mathcal{H}}
\newcommand{\Linvad}{\tilde{\calL}_{t}^{+}}
\newcommand{\Linv}{\calL_{t}^{+}}
\newcommand{\Hinvad}{\tilde{\calH}_{t}^{+}}
\newcommand{\Hinv}{\calH_{t}^{+}}

\newcommand{\Dinvad}{\tilde{\calD}_{t}^{+}}

\newcommand{\gamd}{\gamma_{\downarrow}}
\newcommand{\gamu}{\gamma_{\uparrow}}

\begin{document}

%\preprint{}

\title{Thermodynamic speed limit for non-adiabatic work and its classical-quantum decomposition}

\author{Aoi Yamauchi}
\affiliation{Department of Applied Physics, The University of Tokyo, 7-3-1 Hongo, Bunkyo-ku, Tokyo 113-8656, Japan}

\author{Rihito Nagase}
\affiliation{Department of Applied Physics, The University of Tokyo, 7-3-1 Hongo, Bunkyo-ku, Tokyo 113-8656, Japan}

\author{Kaixin Li}
\affiliation{Department of Applied Physics, The University of Tokyo, 7-3-1 Hongo, Bunkyo-ku, Tokyo 113-8656, Japan}

\author{Takahiro Sagawa}
\affiliation{Department of Applied Physics, The University of Tokyo, 7-3-1 Hongo, Bunkyo-ku, Tokyo 113-8656, Japan}
\affiliation{Quantum-Phase Electronics Center (QPEC), The University of Tokyo, 7-3-1 Hongo, Bunkyo-ku, Tokyo 113-8656, Japan}

\author{Ken Funo}
\affiliation{Department of Applied Physics, The University of Tokyo, 7-3-1 Hongo, Bunkyo-ku, Tokyo 113-8656, Japan}
\email{funo@ap.t.u-tokyo.ac.jp}

\date{\today}

\begin{abstract}
Understanding the fundamental constraint on work far beyond the adiabatic regime is crucial to investigating fast and efficient energy extraction or consumption processes.
In this study, we derive thermodynamic speed limits for non-adiabatic work and quantify the fundamental costs of non-adiabatic work extraction or consumption processes in open quantum systems, where the costs are quantified by geometric and thermodynamic quantities.  
We further decompose the non-adiabatic work into classical and quantum contributions and derive their thermodynamic speed limits, clarifying the classical and quantum nature of the fundamental costs. The obtained results are numerically demonstrated by driven two-level systems. 
\end{abstract}

\maketitle

\section{Introduction}
Quantifying the universal bound on mechanical work is indispensable for optimization of efficient energy conversion, extraction, and storage in various thermodynamic devices, including molecular machines~\cite{seifert2012stochastic, leighton2024flowenergyinformationmolecular}, heat engines~\cite{seifert2012stochastic, delCampo2018, 10.1116/5.0083192, CANGEMI20241}, information engines~\cite{parrondo2015thermodynamics}, and quantum batteries~\cite{RevModPhys.96.031001}. 
According to the second law of thermodynamics, the work cost is minimized by adiabatically slow processes in which the system is always at equilibrium. However, realistic systems typically operate beyond the adiabatic regime and are out of equilibrium; therefore, recent studies aim to develop finite-time and nonequilibrium thermodynamic relations~\cite{SeifertTUR, horowitz2020thermodynamic, doi:10.1073/pnas.2321112121} and optimal protocols~\cite{e19040136, Deffner_2020, Guéry-Odelin_2023, Blaber_2023}. 
When the system operates quasi-adiabatically, one could utilize linear-response relations and thermodynamic metrics to explore geometric bounds and optimization methods for work in both classical and quantum systems~\cite{PhysRevLett.124.040602, Scandi-Thermodynamic-2019, e22101076, PhysRevLett.123.230603, PhysRevLett.131.210401}. 

Far beyond the adiabatic regime, there exist several frameworks, including quantum and thermodynamic speed limits~\cite{deffner2017quantum, Gong-Hamazaki-review, PhysRevX.6.021031, Shiraishi-Speed-2018, Funo-Speed-2019, Nicholson-Time-information-2020, PhysRevX.12.011038, Ito-Stochastic-2020, PhysRevLett.126.010601, PhysRevX.13.011013,   sekiguchi2024}, thermodynamic uncertainty relations~\cite{barato2015thermodynamic,gingrich2016dissipation,horowitz2017proof,horowitz2020thermodynamic}, finite-time Landauer principles~\cite{goold2015nonequilibrium,proesmans2020finite,zhen2021universal,van2022finite} and thermodynamic trade-off relations~\cite{Gingrich-Dissipation-2016, Shiraishi-Universal-2016, Tajima-Superconducting-like-2021, PhysRevResearch.7.013244}, which give universal bounds on the speed of probability currents and current-like observables. For such current-like thermodynamic quantities, it is also known that information-geometric quantities such as Fisher information~\cite{cover1999elements} provide important thermodynamic trade-off relations~\cite{dechant2018multidimensional,hasegawa2019uncertainty,ito2020stochastic,otsubo2020estimating}.
These relations clarify that larger entropy production is required to increase the speed of physical processes when the system is subject to thermal environments. However, existing speed limits and trade-off relations cannot be directly applied to work, since work is not a current-like time-independent observable. While some studies bound the amount of work based on internal energy and heat~\cite{Aghion_2023}, or by utilizing optimal transport theory~\cite{Benamou_Brenier} for classical Langevin systems~\cite{PhysRevLett.106.250601, PhysRevLett.125.100602}, quantification of thermodynamic costs that constrain the speed of non-adiabatic work extraction or consumption processes, particularly in the quantum regime, has not been well established. 

In this study, we derive thermodynamic speed limits for non-adiabatic work in open quantum systems and clarify the fundamental costs of generating work at arbitrary speed. The costs are quantified by the entropy production and the quantum Fisher information metric, allowing thermodynamic and geometric interpretation of the speed limit. We further consider the classical-quantum decomposition of non-adiabatic work and derive their speed limits to characterize the classical and quantum nature of the fundamental costs. This decomposition is particularly meaningful when the drive is purely classical or quantum, as demonstrated by a two-level system example.

This paper is organized as follows. In Sec.~\ref{sec:setup}, we explain the setup of our paper. 
In Sec.~\ref{sec:result}, we derive our first main result, the thermodynamic speed limit for non-adiabatic work. In Sec.~\ref{sec:main2}, we derive our second main result, the classical-quantum decomposition of the non-adiabatic work and their thermodynamic speed limits. In Sec.~\ref{sec:special}, we discuss several limiting cases of the obtained results. In Sec.~\ref{sec:example}, we consider driven two-level systems as an example. We summarize our results in Sec.~\ref{sec:conclusion}.  

\section{\label{sec:setup}Setup}
We assume that the quantum system of interest is weakly coupled to a heat bath at an inverse temperature $\beta$. By assuming the standard Born-Markov-Secular approximations, we arrive at the Markovian master equation of the Gorini-Kossakowski-Sudarshan-Lindblad form~\cite{GKS, Lindblad, Breuer}, given by
\begin{equation}
\dot{\rho}_{t}=\mathcal{L}_t[\rho_t]=\mathcal{H}_t[\rho_t]+\mathcal{D}_{t}[\rho_t], \label{ME}
\end{equation}
where $\dot{X}_{t}:=\partial_{t}X_{t}$ is a short-hand notation of the time-derivative of the quantity $X_{t}$, and 
\beq
\mathcal{H}_{t}[\rho_{t}]:=-\frac{i}{\hbar}[H_t,\rho_t] \label{Hcommutator}
\eeq
describes the effect of unitary time-evolution generated by the time-dependent Hamiltonian $H_{t}=H[\lambda_{t}]$. Here, we denote the time-dependence of the Hamiltonian by a function $\lambda_{t}$. The second term on the right-hand side of Eq.~(\ref{ME}) is the dissipator
\begin{equation}
    \mathcal{D}_{t}[\rho_t]:=\sum_{\omega_t}\gamma(\omega_t)\left(L_{\omega_t}\rho_t L_{\omega_t}^{\dag}-\frac{1}{2}\left\{L_{\omega_t}^{\dag}L_{\omega_t},\rho_t\right\}\right), \label{Dissipator}
\end{equation}
which describes the effect of heat bath, where $\gamma(\omega_{t})$ is the rate and $L_{\omega_{t}}$ is the jump operator that allows the system to jump from one energy eigenstate to another with their energy difference equal to $\hbar \omega_{t}$, and satisfies $[L_{\omega_t},H_t]=\hbar\omega_t L_{\omega_t}$. We further assume the detailed balance condition 
\begin{equation}
    \frac{\gamma(\omega_t)}{\gamma(-\omega_t)}=\exp(\beta\hbar\omega_t)
\end{equation}
to make Eq.~(\ref{ME}) thermodynamically consistent. In particular, the instantaneous steady-state of the system is given by the thermal distribution $\rho^{\mathrm{eq}}_{t}:=\exp(-\beta H_{t})/Z_{t}$ with $Z_{t}:=\Tr[\exp(-\beta H_{t})]$, that is
\begin{equation}
\mathcal{L}_t[\rho^{\mathrm{eq}}_t]=0.
\end{equation}
During time evolution, work is invested in the system from the external control that drives the Hamiltonian, defined as 
\beq
\dot{W}:=\Tr[ \rho_{t} \dot{H}_{t}].
\eeq
In the adiabatically slow driving limit, the density operator of the system can be approximated by its instantaneous thermal state $\rho(t)= \rho_{t}^{\mathrm{eq}}+O(\dot{\lambda}_{t})$. In this regime, the work reduces to the adiabatic work, defined as 
\beq
\dot{W}_{\rm ad} := \Tr[ \dot{H}_{t}\rho_{t}^{\mathrm{eq}}]=\dot{F}_{\rm eq}, \label{Wad}
\eeq
where $F_{\rm eq}:=-\beta^{-1}\ln Z_{t}$ is the equilibrium free energy. When driving is quasi-adiabatic, one can use the adiabatic perturbation theory and obtain the next-order correction to Eq.~(\ref{Wad}), expressed by
\beq
\dot{W}=\dot{W}_{\rm ad} + \Tr[\dot{H}_{t}\mathcal{L}_{t}^{+}[\dot{\rho}_{t}^{\mathrm{eq}}]] + O(\dot{\lambda}_{t}^{3}),  \label{Wlinear}
\eeq
where $\mathcal{L}_{t}^{+}$ is the Drazin inverse of the Liouville superoperator $\mathcal{L}_{t}$, defined as~\cite{Drazin-Pseudo-inverses-1958, Scandi-Thermodynamic-2019, PhysRevE.104.034117} 
\begin{equation}
    \mathcal{L}^+_t[B]:= \int_0^{\infty}d\nu\, e^{\nu\mathcal{L}_t}(\rho^{\mathrm{eq}}_t\Tr[B]-B).
\end{equation}
If $\mathcal{L}_{t}$ is diagonalizable, let us denote it as $\mathcal{L}_{t}=\sum_{\Gamma}\Gamma P_{\Gamma}$, where $\Gamma$ and $P_{\Gamma}$ are the eigenvalues and projection superoperators, respectively. Then, the Drazin inverse $\Linv$ can be expressed as $\Linv=\sum_{\Gamma\neq 0}\Gamma^{-1}P_{\Gamma}$, where \revision{the real part of $-\Gamma^{-1}$ quantifies the relaxation time-scales for each eigenmode.}  

It should be noted that the second term on the right-hand side of Eq.~(\ref{Wlinear}) can be expressed as a thermodynamic metric in the control parameter space $\lambda_{t}$, and it is possible to optimize the work by finding the optimal driving protocol $\lambda_{t}$~\cite{Scandi-Thermodynamic-2019}. 
However, when driving is beyond this near-adiabatic regime, we need a general framework that quantifies the non-adiabatic contribution to the work.
To this end, we employ techniques developed in the study of thermodynamic speed limits and derive a general upper bound on the non-adiabatic work $\dot{W}-\dot{W}_{\rm ad}$. 

\section{\label{sec:result}Main result 1: thermodynamic speed limits for non-adiabatic work}

In this section, we derive a general upper bound on the non-adiabatic work. To begin with, we use the following property of the Drazin inverse (see Appendix~\ref{appendix:Drazin}): 
\beq
\rho_{t}-\rho_{t}^{\mathrm{eq}}=\Linv[\calL_{t}[\rho_{t}]]=\Linv[\dot{\rho}_{t}], 
\eeq
and express the non-adiabatic work as
\beq
\dot{W}-\dot{W}_{\rm ad} = \Tr[\dot{H}_{t}\mathcal{L}_{t}^{+}[\dot{\rho}_{t}]] = \Tr[\Linvad[\dot{H}_{t}]\dot{\rho}_{t}] \label{nonadW}.
\eeq
Here, $\tilde{\mathcal{K}}$ denotes the adjoint of the superoperator $\mathcal{K}$, defined by the relation $\langle X,\mathcal{K}[Y] \rangle=\langle \tilde{\mathcal{K}}[X],Y\rangle$, with $\langle X,Y\rangle:=\Tr[X^{\dagger}Y]$ being the Hilbert-Schmidt scalar product. 
The explicit form of  $\tilde{\mathcal{L}}_{t}^{+}$ reads
\beq
\tilde{\mathcal{L}}_{t}^{+}[B] := \int^{\infty}_{0}d\nu e^{\nu \tilde{\calL}_{t}}\Bigl( \Tr[\rho_{t}^{\mathrm{eq}}B]I-B \Bigr), \label{def_Linvad}
\eeq
where $I$ is the identity operator and $\tilde{\calL}_{t}$ is the adjoint of the Lindblad superoperator defined by 
\beq
\tilde{\mathcal{L}}_{t}(B) := i[H,B] + \sum_{\omega_{t}}\gamma(\omega_{t})\Bigl( L_{\omega_{t}}^{\dagger}B L_{\omega_{t}} - \frac{1}{2}\{ L_{\omega_{t}}^{\dagger}L_{\omega_{t}},B\}\Bigr). \label{sup_Ladjoint}
\eeq 
It should be noted that unlike Eq.~(\ref{Wlinear}), Eq.~(\ref{nonadW}) is exact for any driving speed and contains all orders of $\dot{\lambda}_{t}$. 

\revision{
Based on the expression~(\ref{nonadW}), we derive a thermodynamic speed limit for non-adiabatic work as
\beq
|\dot{W}-\dot{W}_{\rm ad}| \leq \underbrace{\sqrt{4g_{tt}^{\rm QF}V_{\rho_t}(\Linvad[\dot{H}_t])}}_{\text{non-diagonal}} + \underbrace{\sqrt{ 2A(\rho_{t})\dot{\sigma}}}_{\text{diagonal}}=: \mathcal{B}_{1}^{\rm nd}+\mathcal{B}_{1}^{\rm d}, \label{mainresult}
\eeq
which is the first main result of this paper. The proof of Eq.~(\ref{mainresult}) is shown in Appendix~\ref{appendix:derivation}, where we utilize the techniques developed in Refs.~\cite{Funo-Speed-2019, sekiguchi2024}. Here, the bound contains two terms $\mathcal{B}_{1}^{\rm nd}$ and $\mathcal{B}_{1}^{\rm d}$, depending on the non-diagonal and diagonal contributions using the basis of the density operator $\rho_{t}$, respectively. 
}
In the following, we explain each term that appears in Eq.~(\ref{mainresult}).  

\begin{itemize}

\item The quantity $g^{\rm QF}_{tt}$ describes a purely quantum contribution to the quantum Fisher information metric and is defined as~\cite{PhysRevLett.72.3439, PhysRevX.6.021031} 
\beq
g^{\rm QF}_{tt} := \frac{1}{2}\sum_{m\neq n}\frac{(p_{n}-p_{m})^{2}}{p_{n}+p_{m}}|\langle n|\partial_{t}m\rangle|^{2}, \label{QFIM}
\eeq
where we denote the spectral decomposition of the density operator as $\rho_{t}=\sum_{n}p_{n}(t)|n(t)\rangle\langle n(t)|$, and $p_{n}(t)$ and $|n(t)\rangle$ are the eigenvalues and eigenstates of $\rho_{t}$, respectively. We also use the notation $\partial_{t}|m(t)\rangle = |\partial_{t}m\rangle$. This quantity characterizes a purely quantum contribution to the quadratic sensitivity of the time-variation of density operators (see Appendix~\ref{appendix:remarkQFIM} for details), and also quantifies the speed of coherent, unitary part of the time-evolution (see Eq.~(\ref{QFI=QFIM})). 

\item The quantity $V_{\rho_{t}}(\Linvad[\dot{H}_t])$ quantifies the fluctuation of $\Linvad[\dot{H}_t]$, where the variance is defined as $V_{\rho_{t}}[X]:=\Tr[X^{2}\rho_{t}]-(\Tr[X\rho_{t}])^{2}$. 
When $\calL_{t}$ is diagonalizable, $\Linvad$ rescales $\dot{H}_{t}$ by dividing it with effective relaxation rates $\Gamma$ for each eigenmode, meaning that $\Linvad[\dot{H}_{t}]$ characterizes the relative time-scale between the driving speed $\dot{\lambda}_{t}$ of the Hamiltonian and the relaxation rates $\Gamma$. The analytical expression of $\Linvad[\dot{H}_t]$ for a two-level system example is given in Eq.~(\ref{LinvadH_ex}).

\item The activity-like quantity $A$ is defined as
\begin{equation}
A(\rho_{t}):=\frac{1}{2}\sum_{n}\langle n|\Linvad[\dot{H}_{t}]|n\rangle^{2}\langle n| \mathcal{D}_{t}^{\rm sym}[\rho_{t}]|n\rangle, \label{defA}
\end{equation}
where
\beq
\calD_{t}^{\rm sym}[\rho_{t}] := \sum_{\omega_{t}}\gamma(\omega_{t}) \Bigl( L_{\omega_{t}}\rho_{t}L_{\omega_{t}}^{\dagger}+\frac{1}{2}\{L_{\omega_{t}}^{\dagger}L_{\omega_{t}},\rho_{t}\} \Bigr)
\eeq
is defined by replacing the minus sign with the plus sign in front of the second term of the dissipator~(\ref{Dissipator}). In particular,
\beq
A_{\rm act}:=\frac{1}{2}\sum_{n}\langle n| \calD_{t}^{\rm sym}[\rho_{t}]|n\rangle=\frac{1}{2}\Tr[\calD_{t}^{\rm sym}[\rho_{t}]]=\sum_{\omega_{t}}\gamma(\omega_{t})\Tr[L_{\omega_{t}}^{\dagger}L_{\omega_{t}}\rho_{t}]
\eeq
quantifies the number of total jumps per unit time and is called (dynamical) activity~\cite{Shiraishi-Speed-2018, Funo-Speed-2019}. Therefore, the term $A$ is an activity-like quantity that quantifies the instantaneous fluctuation of $\Linvad[\dot{H}_{t}]$ multiplied by the time-scale of quantum jumps for each eigenbasis $|n(t)\rangle$ of $\rho_{t}$. \revision{See also Eq.~(\ref{Adiag_ex}) for the simplified expression of $A$ in the case of two-level systems.}

\item The entropy production rate $\dot{\sigma}$ is defined as 
\beq
\dot{\sigma}:=\dot{S}-\beta \dot{Q} \geq 0, \label{EP}
\eeq
where $S(\rho):=-\Tr[\rho\ln\rho]$ is the von Neumann entropy of the system, $\dot{Q}:=\Tr[ H_{t}\dot{\rho}_{t}]$ is the heat that quantifies the energy exchange between the system and the heat bath, and the nonnegativity of $\dot{\sigma}$ represents the second law of thermodynamics~\cite{10.1063/1.523789, Funo2018}. The entropy production quantifies the thermodynamic cost of non-adiabatic processes and vanishes in the adiabatic limit. 

\end{itemize}

The obtained thermodynamic speed limit~(\ref{mainresult}) is valid for any driving speed and represents a general upper bound on non-adiabatic work. 
\revision{
In the adiabatic limit, both $g_{tt}^{\rm QF}$ and $\dot{\sigma}$ vanish, thereby the bound quantifies geometric and thermodynamic costs to produce work in the non-adiabatic regime. 

The non-diagonal part of the bound $\mathcal{B}_{1}^{\rm nd}$ depends on $g_{tt}^{\rm QF}$, which characterizes the speed of coherent unitary part of the time-evolution. Moreover, it vanishes in the limit where Eq.~(\ref{ME}) can be approximated by a classical master equation. On the other hand, $\mathcal{B}_{1}^{\rm d}$ depends on the entropy production $\dot{\sigma}$ and the activity-like quantity $A$, which quantifies the speed of the stochastic jump part of the time-evolution~\cite{Shiraishi-Speed-2018, Funo-Speed-2019}. Moreover, it vanishes in the limit where Eq.~(\ref{ME}) can be approximated by the time-evolution for isolated quantum systems.
Based on the above discussion, the bounds $\mathcal{B}_{1}^{\rm nd}$ and $\mathcal{B}_{1}^{\rm d}$ can be interpreted as the quantum and classical part of the speed limit for non-adiabatic work, respectively.} In the next section, we generalize this classical-quantum interpretation by deriving another bound based on the classical-quantum decomposition of the non-adiabatic work.

\section{\label{sec:main2}Main result 2: Classical-quantum decomposition of the non-adiabatic work and their thermodynamic speed limits}

In this section, we decompose the non-adiabatic work into classical and quantum contributions and derive thermodynamic speed limits for individual terms. \revision{Specifically, instead of splitting the bound into diagonal and non-diagonal contributions using the basis of $\rho_{t}$ as done in Sec.~\ref{sec:result}, we now use the basis of $H_{t}$ and derive a similar bound.}

\subsection{Classical-quantum decomposition of the non-adiabatic work}
\revision{To begin with, we decompose the time derivative of the Hamiltonian into the classical and quantum contributions, defined by
\beq
\dot{H}_{t}^{\rm cl} := \sum_{n}\dot{\epsilon}_{n}|\epsilon_{n}\rangle\langle \epsilon_{n}|, \ \ 
\dot{H}_{t}^{\rm qm} := \sum_{n}\epsilon_{n} (|\dot{\epsilon}_{n}\rangle\langle \epsilon_{n}| + | \epsilon_{n}\rangle\langle \dot{\epsilon}_{n}| ). \label{Hamiltonian_decomposition}
\eeq
Based on the decomposition~(\ref{Hamiltonian_decomposition}), the non-adiabatic work can be similarly decomposed into classical and quantum contributions such as} $\dot{W}=\dot{W}_{\rm cl}+\dot{W}_{\rm qm}$, where
\beq
\dot{W}_{\rm cl}:=\sum_{n}\dot{\epsilon}_{n}\langle \epsilon_{n}|\rho_{t}|\epsilon_{n}\rangle=\Tr[\dot{H}_{t}^{\rm cl}\rho_{t}^{\rm cl}] \label{defWcl}
\eeq
is the work originating from the time variation of the energy eigenvalues $\epsilon_{n}(t)$, and depends only on the diagonal component of the density operator using the energy eigenbasis, i.e., $\rho_{t}^{\rm cl}=\sum_{n}|\epsilon_{n}\rangle\langle\epsilon_{n}|\rho_{t}|\epsilon_{n}\rangle\langle\epsilon_{n}|$. Therefore, Eq.~(\ref{defWcl}) quantifies the classical contribution to the work. On the other hand,
\beq
\dot{W}_{\rm qm}:=\sum_{n}\epsilon_{n}\Bigl(\langle \dot{\epsilon}_{n}|\rho_{t}|\epsilon_{n}\rangle+\langle \epsilon_{n}|\rho_{t}|\dot{\epsilon}_{n}\rangle\Bigr)=\Tr[\dot{H}_{t}^{\rm qm}\rho_{t}^{\rm qm}] \label{defWqm}
\eeq
depends solely on the coherence of the density operator $\rho_{t}^{\rm qm}:=\rho_{t}-\rho_{t}^{\rm cl}$ and thus quantifies quantum contribution to the work. 
We note that in the adiabatic limit, we have $\dot{W}_{\rm cl}=\dot{W}_{\rm ad}+O(\dot{\lambda}_{t}^{2})$ and $\dot{W}_{\rm qm}=O(\dot{\lambda}_{t}^{2})$. Therefore, the non-adiabatic contribution to the work is decomposed as
\beq
\dot{W}-\dot{W}_{\rm ad} = \underbrace{( \dot{W}_{\rm cl}-\dot{W}_{\rm ad} )}_{\text{classical}} + \underbrace{\dot{W}_{\rm qm}}_{\text{quantum}}. \label{Wdecomp2}
\eeq
In the following, we discuss how each term in Eq.~(\ref{Wdecomp2}) can be bounded. 

\subsection{Thermodynamic speed limits for non-adiabatic classical work}
To derive an upper bound on the non-adiabatic classical work, we start from the expression
\revision{
\beq
\dot{W}_{\rm cl}-\dot{W}_{\rm ad} = \Tr[\dot{H}_{t}^{\rm cl}(\rho_{t}^{\rm cl}-\rho_{t}^{\rm eq}) ] = \Tr[ \dot{H}_{t}^{\rm cl}\calD_{t}^{+}[\calD_{t}[\rho_{t}^{\rm cl}]]], \label{WDinv}
\eeq
where $\calD_{t}^{+}$ is the Drazin inverse of $\calD_{t}$, and we have used the relation $\rho_{t}^{\rm cl}-\rho_{t}^{\rm eq}=\calD_{t}^{+}[\calD_{t}[\rho_{t}^{\rm cl}]]$, which follows from Eq.~(\ref{appendix:propertyD}).}
By following a derivation similar to that for Eq.~(\ref{mainresult}), which is detailed in Appendix~\ref{appendix:secondproof}, the non-adiabatic classical work $\dot{W}_{\rm cl}-\dot{W}_{\rm ad}$ is bounded as
\beq
|\dot{W}_{\rm cl}-\dot{W}_{\rm ad}| \leq \sqrt{2A(\rho_{t}^{\rm cl})\dot{\sigma}_{\rm cl}} =: \mathcal{B}_{2}^{\rm cl} , \label{Wclbound}
\eeq
where 
\beq
\dot{\sigma}_{\rm cl} := \dot{S}_{\rm cl}-\beta\dot{Q} \geq 0
\eeq
quantifies the incoherent part of the entropy production, with $S_{\rm cl}:=S(\rho_{t}^{\rm cl})$ being the diagonal entropy. We can further show that~\cite{santos2019role, RevModPhys.93.035008, Tajima-Superconducting-like-2021} (see Eq.~(\ref{sigma-sigmacl}))
\beq
\dot{\sigma}_{\rm cl}\leq \dot{\sigma}, \label{sigmaclbd}
\eeq
which means that the generation of coherence increases the entropy production. The explicit expression of $A$ in Eq.~(\ref{Wclbound}) reads
\beq
A(\rho_{t}^{\rm cl})=\frac{1}{2}\sum_{n}\langle \epsilon_{n}|\Dinvad[\dot{H}_{t}^{\rm cl}]|\epsilon_{n}\rangle^{2}\langle \epsilon_{n}| \mathcal{D}_{t}^{\rm sym}[\rho_{t}^{\rm cl}]|\epsilon_{n}\rangle. 
\eeq
This expression follows from the definition of $A(\rho_{t})$ in Eq.~(\ref{defA}), by noting that the diagonal basis of $\rho_{t}^{\rm cl}$ is given by $|\epsilon_{n}\rangle$, and that $\langle \epsilon_{n}|\Linvad[\dot{H}_{t}]|\epsilon_{n}\rangle=\langle\epsilon_{n}|\Dinvad[\dot{H}_{t}^{\rm cl}]|\epsilon_{n}\rangle$, where $\Dinvad$ is the adjoint of $\calD_{t}^{+}$. 

The obtained bound~(\ref{Wclbound}) shows that the non-adiabatic classical work is bounded by the incoherent part of the entropy production and the activity-like quantity. 
Therefore, the bound~(\ref{Wclbound}) characterizes the incoherent, classical part of the thermodynamic speed limit on non-adiabatic work and vanishes when the dynamics is purely coherent (see also Sec.~\ref{sec:drive:quantum}). 

\subsection{Thermodynamic speed limits for non-adiabatic quantum work}
Next, we derive an upper bound on the non-adiabatic quantum work by using the expression
\revision{
\beq
\dot{W}_{\rm qm}=\Tr[\dot{H}_{t}^{\rm qm}\rho_{t}^{\rm qm}]=\Tr[ \dot{H}_{t}^{\rm qm}\calH_{t}^{+}[\calH_{t}[\rho_{t}]]], \label{WqmHinv}
\eeq
where $\calH_{t}^{+}$ is the Drazin inverse of $\calH_{t}$ defined in Eq.~(\ref{Hinvad}), and we have used the relation $\rho_{t}^{\rm qm}=\calH_{t}^{+}[\calH_{t}[\rho_{t}]]$, which follows from Eq.~(\ref{appendix:H}).} Now, $\dot{W}_{\rm qm}$ can be bounded as (see Appendix~\ref{appendix:secondproof} for the proof):
\beq
|\dot{W}_{\rm qm}| \leq \frac{1}{\hbar}\sqrt{\mathcal{F}_{\rho_{t}}(H_{t})V_{\rho_{t}}(H_{\rm cd}) } =: \mathcal{B}_{2}^{\rm qm}, \label{Wqmbound}
\eeq
where 
\beq
    \mathcal{F}_{\rho_{t}}(H_{t}):=2\sum_{m\neq n}\frac{(p_n-p_m)^2}{p_n+p_m}\left|\langle n|H_{t}|m\rangle \right|^2
\eeq
is the quantum Fisher information that quantifies the coherence of the density operator $\rho_{t}^{\rm qm}$, and 
\beq
H_{\rm cd}:=i\hbar \sum_{n}(I-|\epsilon_{n}\rangle\langle \epsilon_{n}|)|\partial_{t}\epsilon_{n}\rangle \langle \epsilon_{n}|
\eeq
is the counter-diabatic Hamiltonian that is used as a control Hamiltonian to guide the system along the instantaneous energy eigenstate~\cite{RevModPhys.91.045001}. \revision{Note that $V_{\rho_{t}}(H_{\rm cd})$ that appears in the bound $\mathcal{B}_{2}^{\rm qm}$ is the energy fluctuation of the control Hamiltonian $H_{\rm cd}$, which quantifies the speed of controlling non-adiabatic time-evolution in isolated~\cite{PhysRevLett.118.100602, PhysRevX.9.011034} and quasi-adiabatic open systems~\cite{Funo-Speed-2019}. Moreover, $V_{\rho_{t}}(H_{\rm cd})$ sets an upper bound on the work fluctuation in isolated systems controlled by $H_{\rm cd}$~\cite{PhysRevLett.118.100602}.}

The obtained bound~(\ref{Wqmbound}) shows that the non-adiabatic quantum work is bounded by the amount of coherence multiplied by the energy variance of $H_{\rm cd}$, where the latter quantifies how the energy eigenstates vary in time via non-adiabatic drives. Therefore, the bound~(\ref{Wqmbound}) characterizes coherent, quantum part of the thermodynamic speed limit on non-adiabatic work and vanishes in the classical limit. 

We further note that from Eq.~(\ref{Hinvad=Hcd}), we have
\beq
\Hinvad[\dot{H}_{t}]=\Hinvad[\dot{H}_{t}^{\rm qm}]=H_{\rm cd},
\eeq
and thus $V_{\rho_t}(\Linvad[\dot{H}_t])$ that appears in Eq.~(\ref{mainresult}) reduces to $V_{\rho_{t}}(H_{\rm cd})$. Similarly, $4g_{tt}^{\rm QF}$ reduces to $\hbar^{-2}\mathcal{F}_{\rho_{t}}(H_{t})$ when the generator is given by $\calH_{t}$ [see Eq.~(\ref{QFI=QFIM})]. Therefore, the first term of the bound in Eq.~(\ref{mainresult}) can also be interpreted as a quantum contribution to the bound of the non-adiabatic work. 

\subsection{Classical-quantum decomposition of the thermodynamic speed limit}

We now use the triangle inequality $|\dot{W}-\dot{W}_{\rm ad}|\leq |\dot{W}_{\rm cl}-\dot{W}_{\rm ad}|+|\dot{W}_{\rm qm}|$ and combine Eqs.~(\ref{Wclbound}) and (\ref{Wqmbound}) to derive another version of the non-adiabatic work bound as
\beq
|\dot{W}-\dot{W}_{\rm ad}|\leq  \underbrace{\frac{1}{\hbar}\sqrt{\mathcal{F}_{\rho_{t}}(H_{t})V_{\rho_{t}}(H_{\rm cd}) }}_{\text{quantum}} + \underbrace{\sqrt{2A(\rho_{t}^{\rm cl})\dot{\sigma}_{\rm cl}}}_{\text{classical}} = \mathcal{B}_{2}^{\rm qm} + \mathcal{B}_{2}^{\rm cl}
 . \label{mainresult2}
\eeq
These classical-quantum decomposition of the thermodynamic speed limits for non-adiabatic work [Eqs.~(\ref{Wclbound}), (\ref{Wqmbound}), and (\ref{mainresult2})] are the second main result of this paper. \revision{We again note that the splitting of the bound into $\mathcal{B}_{2}^{\rm qm}$ and $\mathcal{B}_{2}^{\rm cl}$ is based on the decomposition with respect to the non-diagonal and diagonal contributions using the basis of the Hamiltonian $H_{t}$, respectively.}

By combining~(\ref{mainresult}) and (\ref{mainresult2}), we obtain  
\beq
|\dot{W}-\dot{W}_{\rm ad}|\leq  \min\left\{ \mathcal{B}_{1}^{\rm nd}+\mathcal{B}_{1}^{\rm d}, \  \mathcal{B}_{2}^{\rm qm}+\mathcal{B}_{2}^{\rm cl}  \right\}, \label{mainresult_combined}
\eeq
where the minimum of the right-hand side of~(\ref{mainresult_combined}) depends on the details of the model.  

\section{\label{sec:special}Special cases of the speed limit}
In this section, we consider two limiting cases of the obtained result.

\subsection{\label{sec:drive:classical}Classical drive}
First, let us consider a classical driving of the Hamiltonian by only changing its energy eigenvalues $\epsilon_{n}(t)$ in time and keeping the eigenstates constant. Then, we have $\dot{H}_{t}=\dot{H}_{\rm cl}$ and $\dot{W}=\dot{W}_{\rm cl}$. Therefore, the second main result~(\ref{mainresult2}) reduces to
\beq
|\dot{W}-\dot{W}_{\rm ad}| \leq  \sqrt{ 2A(\rho_{t})\dot{\sigma}} \hspace{5mm} (\text{classical drive}).  \label{mainresult_cl}
\eeq

It should also be noted that when the initial density operator satisfies $\rho_{0}=\rho^{\rm cl}_{0}$, we have $\rho_{t}=\rho^{\rm cl}_{t}$ at later times $t$. By assuming this initial condition, we have $g_{tt}^{\rm QF}=0$ and $\dot{\sigma}_{\rm cl}=\dot{\sigma}$. Note that this setup is essentially equivalent to considering classical stochastic systems, and the first main result~(\ref{mainresult}) also reduces to Eq.~(\ref{mainresult_cl}). 

\subsection{\label{sec:drive:quantum}Quantum drive}
Next, let us consider the opposite situation compared to Sec.~\ref{sec:drive:classical}, i.e., the energy eigenstates are time-dependent $|\epsilon_{n}(t)\rangle$ but the energy eigenvalues $\epsilon_{n}$ do not depend on time. For this quantum drive, we have $\dot{H}_{t}=\dot{H}_{t}^{\rm qm}$, $\dot{W}=\dot{W}^{\rm qm}$, and $\dot{W}_{\rm ad}=0$. 
Therefore, the second main result~(\ref{Wqmbound}) reduces to
\beq
|\dot{W}-\dot{W}_{\rm ad}| \leq  \frac{1}{\hbar}\sqrt{\mathcal{F}_{\rho_{t}}(H_{t})V_{\rho_t}(H_{\rm cd})} \hspace{5mm} (\text{quantum drive}).  \label{mainresult_qa}
\eeq

%{\color{red}
%\subsection{quasi-adiabatic regime}

%\beq
%\mathcal{F}_{\rho}(H)=2\sum_{m\neq n}\frac{(\epsilon_{n}-\epsilon_{m})^{2}}{p_{n}^{\rm eq}+p_{m}^{\rm eq}} |\langle \epsilon_{n}|\Linv[\dot{\rho}_{t}^{\rm eq}]|\epsilon_{m}\rangle|^{2}.
%\eeq

%\beq
%V_{\rho}(H_{\rm cd})=\hbar^{2}\sum_{n}p_{n}^{\rm eq} g^{\rm FS}_{tt}(|\epsilon_{n}\rangle),
%\eeq
%where 
%\beq
%g^{\rm FS}_{tt}(|\epsilon_{n}\rangle)=\langle \partial_{t}\epsilon_{n}|(1-|\epsilon_{n}\rangle\langle \epsilon_{n}|)|\partial_{t}\epsilon_{n}\rangle 
%\eeq
%is the Fubini-Study metric for the $n$-th energy eigenstate.
%}

\section{\label{sec:example}Example: two-level system}

\begin{figure}[tbp]
    \centering
    \includegraphics[width=0.9\linewidth]{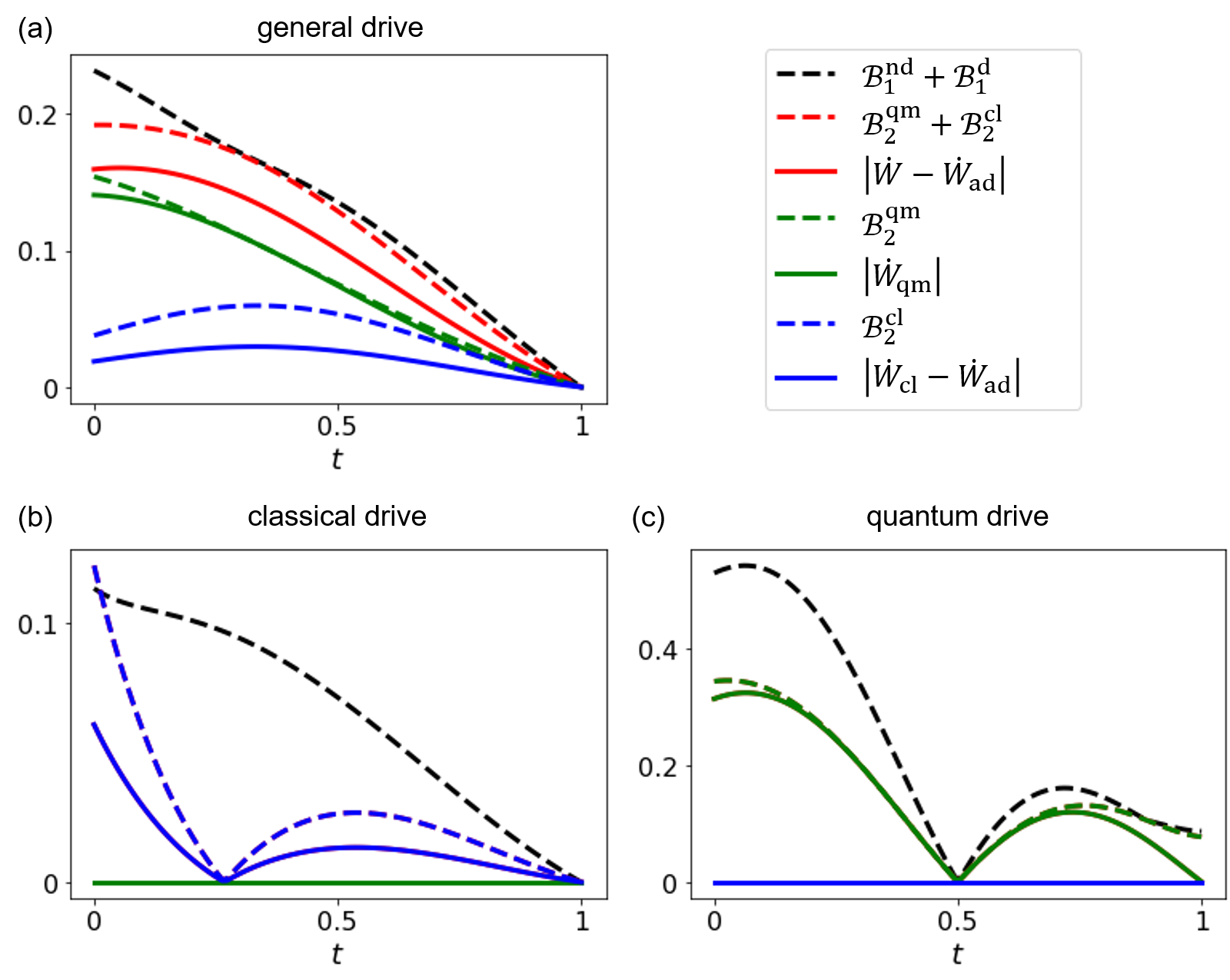}
    \caption{Numerical calculation of the thermodynamic speed limits for non-adiabatic work for two-level systems. Red, green, and blue solid curves plot the non-adiabatic works $|\dot{W}-\dot{W}_{\rm ad}|$, $|\dot{W}_{\rm qm}|$, and $|\dot{W}_{\rm cl}-\dot{W}_{\rm ad}|$, respectively. Black and red dashed curves plot the non-adiabatic work bounds $\mathcal{B}_{1}^{\rm nd}+\mathcal{B}_{1}^{\rm d}$~(\ref{mainresult}) and $\mathcal{B}_{2}^{\rm qm}+\mathcal{B}_{2}^{\rm cl}$~(\ref{mainresult2}), respectively. Blue and green dashed curves are the classical and quantum bounds $\mathcal{B}_{2}^{\rm cl}$~(\ref{Wclbound}) and $\mathcal{B}_{2}^{\rm qm}$~(\ref{Wqmbound}), respectively. (a) General drive. Both classical and quantum terms contribute to the non-adiabatic work and the thermodynamic speed limit~(\ref{mainresult2}). (b) Classical drive $\dot{H}_{t}=\dot{H}_{t}^{\rm cl}$. Note that quantum contributions vanish and the red and blue curves become identical (see also Sec.~\ref{sec:drive:classical}). (c) Quantum drive $\dot{H}_{t}=\dot{H}_{t}^{\rm qm}$. Note that classical contributions vanish and the red and green curves become identical (see also Sec.~\ref{sec:drive:quantum}). The parameters are $\gamma_{0}=\beta=1$, $ p_{+}(0)=0.3,\ \rho_{+-}(0)=0.2+0.1i$. }
    \label{fig:bound}
\end{figure}

In this section, we consider a two-level system and numerically demonstrate the main results. In the following, we set $\hbar=1$ for simplicity. See also Appendix~\ref{appendix:example} for further details. 

We consider the following type of two-level system Hamiltonian:
\begin{equation}
    H_{t}=\frac{\Delta_{t}}{2}\sigma_x+\frac{q_{t}}{2}\sigma_z, \label{example:H}
\end{equation}
where $\Delta_{t}$ and $q_{t}$ denote the time-dependence of the external driving field ($\lambda_{t}=(\Delta_{t},q_{t})$) and $\sigma_{x}$, $\sigma_{z}$ are the Pauli-X-operator and Pauli-Z-operator, respectively. The energy eigenvalues are given by $\epsilon_{\pm}=\pm\omega_{t}/2$, where we define the energy difference as $\omega_{t}:=\sqrt{\Delta_{t}^{2}+q_{t}^{2}}$. 
%We also denote their corresponding energy eigenstates by $|\epsilon_{\pm}\rangle$. 
The energy eigenstates are given by $|\epsilon_{+}\rangle=\cos\theta_{t}|e\rangle + \sin\theta_{t}|g\rangle$ and $|\epsilon_{-}\rangle=\sin\theta_{t}|e\rangle-\cos\theta_{t}|g\rangle$, where $\theta_{t}:=(1/2)\text{cot}^{-1}(q_{t}/\Delta_{t})$, and $|e\rangle$ and $|g\rangle$ are the eigenstates of $\sigma_{z}$, i.e., $\sigma_{z}|e\rangle=|e\rangle$ and $\sigma_{z}|g\rangle=-|g\rangle$.
The time-dependent Hamiltonian~(\ref{example:H}) can be realized, for example, in the charge qubit configuration, where $q_{t}$ can be varied by changing the applied magnetic flux, while $\Delta_{t}$ can be varied by changing the applied gate voltage~\cite{PhysRevB.100.085405}.  
%in various information platforms, including superconducting qubits~\cite{10.1063/1.5089550}, NMR systems~\cite{RevModPhys.76.1037}, and NV-center spins~\cite{annurev:/content/journals/10.1146/annurev-conmatphys-030212-184238}, by applying single-qubit rotation gates. 

The time-evolution equation for the two-level system is assumed to take the following form
\begin{equation}
    \dot{\rho}_{t}=-i[H_t,\rho_t]+\gamma_{\downarrow}\left(\sigma_{t}^-\rho\sigma_{t}^+-\frac{1}{2}\left\{\sigma_{t}^+\sigma_{t}^-,\rho_t\right\}\right)+\gamma_{\uparrow}\left(\sigma_{t}^+\rho\sigma_{t}^--\frac{1}{2}\left\{\sigma_{t}^-\sigma_{t}^+,\rho_t\right\}\right),
\end{equation}
where $\sigma_t^+=\ketbra{\epsilon_+}{\epsilon_-}$ and $    \sigma_t^-=\ketbra{\epsilon_-}{\epsilon_+}$ are the raising and lowering operators between energy eigenstates, and $\gamma_{\downarrow}:=\gamma_{0}(N(\omega_{t})+1)$ and $\gamma_{\uparrow}:=\gamma_{0}N(\omega_{t})$ with $N(\omega_{t}):=1/(e^{\beta\omega_{t}}-1)$ are the transition rates satisfying the detailed balance condition $\gamma_{\downarrow}/\gamma_{\uparrow}=e^{\beta\omega_{t}}$. \revision{In the following, we use the notation $p_{\pm}=\langle \epsilon_{\pm}|\rho|\epsilon_{\pm}\rangle$ and $\rho_{+-}=\langle \epsilon_{+}|\rho|\epsilon_{-}\rangle$,

In this setup, we can derive a simplified analytical expression for the obtained bound~(\ref{mainresult2}). In particular, the counter-diabatic Hamiltonian is given by
    \beq
    H_{\rm cd}= i\dot{\theta}_{t}(|\epsilon_{+}\rangle\langle \epsilon_{-}|-|\epsilon_{-}\rangle\langle\epsilon_{+}|)=\dot{\theta}_{t}\sigma_{y},
    \eeq
and its variance reads
   \beq
    V_{\rho_{t}}(H_{\rm cd})=\dot{\theta}^{2}_{t}\left(1-4(\Im[\rho_{+-}])^{2}\right).
    \eeq
In addition, the activity-like quantity reads
    \beq
    A(\rho_{t}^{\rm cl}) = 2\left(\frac{\dot{\omega}_{t}}{\gamd+\gamu}\right)^{2}A_{\rm act}, \label{Adiag_ex}
    \eeq
    where
    \beq
A_{\rm act}=\gamma_{\uparrow}p_{-}+\gamma_{\downarrow}p_{+}
\eeq
is the activity. This expression demonstrates that $A(\rho_{t}^{\rm cl})$ is essentially given by the total number of jump rates $A_{\rm act}$ multiplied by the square of the relative timescales between the change of the Hamiltonian $\dot{\omega}_{t}$ and the relaxation rates $\gamma_{\uparrow}+\gamma_{\downarrow}$.
}

We now numerically plot the main results~(\ref{mainresult}), (\ref{Wclbound}), (\ref{Wqmbound}), and (\ref{mainresult2}) in Fig.~\ref{fig:bound} by choosing the following three different options for $q_{t}$ and $\Delta_{t}$: 

\begin{description}
\item[(a) General drive] We choose $q_{t}=(1/2)(1+\sin(\pi t/2))$ and $\Delta_{t}=1$. In this case, both classical and quantum terms contribute to the non-adiabatic work and the  thermodynamic speed limit~(\ref{mainresult2}), as shown in Fig.~\ref{fig:bound} (a). 

\item[(b) Classical drive] We choose $q_{t}=(1/2)(1+\sin(\pi t/2))$ and $\Delta_{t}=0$, such that the eigenstates become time-independent $|\epsilon_{+}\rangle=|e\rangle$ and $|\epsilon_{-}\rangle=|g\rangle$. Then, the drive becomes classical $\dot{H}_{t}=\dot{H}_{t}^{\rm cl}$. As discussed in Sec.~\ref{sec:drive:classical}, only the classical term contributes to the non-adiabatic work and the thermodynamic speed limit, which is shown in Fig.~\ref{fig:bound} (b). 

\item[(c) Quantum drive] We choose $q_{t}=\sin(\pi t/2)$ and $\Delta_{t}=\cos(\pi t/2)$, such that the eigenenergies are time-independent $\dot{\omega}_{t}=0$. Then, the drive becomes quantum $\dot{H}_{t}=\dot{H}_{t}^{\rm qm}$. As discussed in Sec.~\ref{sec:drive:quantum}, only the quantum term contributes to the non-adiabatic work and the thermodynamic speed limit, which is shown in Fig.~\ref{fig:bound} (c).
\end{description}

\section{\label{sec:conclusion}Conclusion}
We have derived the thermodynamic speed limit inequalities for non-adiabatic work~(\ref{mainresult}), (\ref{mainresult2}). The obtained results clarify the fundamental costs of work extraction for arbitrary driving speed. We have further introduced the classical-quantum decomposition of the non-adiabatic work and clarified how the fundamental costs are divided into classical and quantum contributions~(\ref{Wclbound}), (\ref{Wqmbound}) [See also Fig.~\ref{fig:bound}]. The classical part of the cost is quantified by the entropy production rate and the activity-like quantity, representing thermodynamic costs to generate stochastic jump dynamics. The quantum part of the cost is quantified by the quantum Fisher information and the variance of the counter-diabatic Hamiltonian, representing geometrical costs to generate coherent unitary dynamics.
 
An interesting future direction is to further investigate the classical-quantum decomposition of non-adiabatic work and analyze the role of quantum coherence in thermodynamics. In the quasi-adiabatic regime, it has been reported in Ref.~\cite{PhysRevLett.123.230603} that generation of quantum coherence violates the fluctuation-dissipation relation. It has also been observed in various models that generation of quantum coherence is typically detrimental to the performance of heat engines~\cite{PhysRevB.100.085405}.
However, a general understanding on the effect of having coherence in the non-adiabatic work extraction processes is not complete and deserves further studies.
Understanding the role of coherence in thermodynamics is anticipated to lead to obtaining the design principle of fast and efficient energy conversion and information processing devices. 

It is also interesting to investigate further the geometric and thermodynamic properties of the obtained bound for nonequilibrium states. Extending the concepts of geometric quantities defined for adiabatic or equilibrium states and their optimization methods~\cite{PhysRevLett.124.040602, Scandi-Thermodynamic-2019, e22101076, PhysRevLett.123.230603, PhysRevLett.131.210401} to nonequilibrium setting is another important future direction. \revision{We also note that the obtained thermodynamic speed limit bounds the absolute value of the non-adiabatic work, while the sign of the non-adiabatic work is shown to be non-negative in the slow-driving regime~\cite{PhysRevLett.123.230603}. It would be interesting if we can further generalize our approach and obtain bounds beyond its absolute value. It would also be an interesting direction if we could derive thermodynamic uncertainty relation bounds on non-adiabatic work, for example, utilizing the fluctuation theorem uncertainty relations~\cite{PhysRevLett.123.090604, PhysRevLett.123.110602}.}

\begin{acknowledgments}
We thank Hiroyasu Tajima for useful discussions. 
This work was supported by JST ERATO Grant No. JPMJER2302, Japan.
R.N. is supported by the World-leading Innovative Graduate Study Program
for Materials Research, Industry, and Technology (MERIT-WINGS) of the University of Tokyo.
T.S. is supported by JSPS KAKENHI Grant No. JP19H05796 and
JST CREST Grant No. JPMJCR20C1. T.S. is also supported
by Institute of AI and Beyond of the University of Tokyo.
K.F. is supported by JSPS KAKENHI Grant Nos. JP23K13036 and JP24H00831. 
\end{acknowledgments}

\appendix

\section{\label{appendix:Drazin}Properties of the Drazin pseudo inverse and its adjoint}

The Drazin inverse of the Lindblad superoperator $\calL^{+}_{t}$ satisfies the following properties:
\begin{equation}
\begin{split}
&\mathcal{L}^+_t\mathcal{L}_t[X]=\mathcal{L}_t\mathcal{L}^+_t[X]=X-\rho^{\mathrm{eq}}_t\Tr[X],\\
    &\mathcal{L}^+_t[\rho^{\mathrm{eq}}_t]=\Tr\left[\mathcal{L}^+_t[X]\right]=0.
\end{split}
\end{equation}
Note that same relations hold for $\calD_{t}^{+}$~(\ref{Dissipator}):  
\begin{equation}
\begin{split}
&\mathcal{D}^+_t\mathcal{D}_t[X]=\mathcal{D}_t\mathcal{D}^+_t[X]=X-\rho^{\mathrm{eq}}_t\Tr[X],\\
&\mathcal{D}^+_t[\rho^{\mathrm{eq}}_t]=\Tr\left[\mathcal{D}^+_t[X]\right]=0. \label{appendix:propertyD}
\end{split}
\end{equation}

%\subsection{\label{appendix:DrazinUnitary}Unitary case}
Next, let us consider the Drazin inverse of $\calH_{t}$~(\ref{Hcommutator}). 
We first note that the steady state of $\calH_{t}$ is no longer unique, since any $\rho_{t}^{\rm cl}$ satisfies $\calH_{t}[\rho_{t}^{\rm cl}]=0$. For given $\rho_{t}$, the steady-state of $\calH_{t}$ reads 
\beq
\lim_{\nu\rightarrow \infty}e^{\nu\calH_{t}}\rho_{t}=\sum_{n}\langle \epsilon_{n}|\rho_{t}|\epsilon_{n}\rangle |\epsilon_{n}\rangle \langle \epsilon_{n}|=\rho_{t}^{\rm cl}, \label{isoSS}
\eeq
where we assume that the Hamiltonian is non-degenerate for simplicity. Therefore, the steady-state of $\calH_{t}$ depends on the operator acting on it. The explicit form of $\Hinv$ reads
\beq
\Hinv[X] = i\hbar \sum_{n\neq m} \frac{|\epsilon_{m}\rangle\langle\epsilon_{m}|X|\epsilon_{n}\rangle\langle\epsilon_{n}|}{\epsilon_{m}-\epsilon_{n}}, \label{Hinvad}
\eeq
and satisfies
\beqa
& &\calH_{t}^{+}\calH_{t}[X]=\calH_{t}\calH_{t}^{+}[X]=X-\sum_{n}\langle \epsilon_{n}|X|\epsilon_{n}\rangle |\epsilon_{n}\rangle \langle \epsilon_{n}|, \label{appendix:H}\\
    & &\Hinv[\rho_{t}^{\rm cl}]=\Tr\left[\Hinv[X]\right]=0.
\eeqa
We also note that 
\beq
\Hinvad[X]=-\Hinv[X].
\eeq
In addition, we have
\beqa
\Hinvad[\dot{H}_{t}] &=& -i\hbar \sum_{n\neq m}\frac{1}{\epsilon_{m}-\epsilon_{n}} \Bigl(\epsilon_{n} |\epsilon_{m}\rangle\langle\epsilon_{m}|\dot{\epsilon}_{n}\rangle\langle\epsilon_{n}| +\epsilon_{m} |\epsilon_{m}\rangle\langle \dot{\epsilon}_{m}|\epsilon_{n}\rangle\langle\epsilon_{n}| \Bigr) \nonumber \\
&=& i\hbar \sum_{n\neq m}  |\epsilon_{m}\rangle\langle\epsilon_{m}|\dot{\epsilon}_{n}\rangle\langle\epsilon_{n}| \nonumber \\
&=& H_{\rm cd}. \label{Hinvad=Hcd}
\eeqa

\section{\label{appendix:derivation}Proof of the first main result}
In this section, we show the first main result~(\ref{mainresult}). To begin with, we use the diagonal-nondiagonal decomposition of the dissipator $ \mathcal{D}[\rho_t]=\mathcal{D}_\mathrm{d}[\rho_t]+\mathcal{D}_\mathrm{nd}[\rho_t]$ introduced in Ref.~\cite{Funo-Speed-2019}, where
\beqa
\mathcal{D}_\mathrm{d}[\rho_t]&:=& \sum_{n}|n\rangle\langle n|\calD_{t}[\rho_{t}]|n\rangle \langle n| , \label{disd} \\
\mathcal{D}_\mathrm{nd}[\rho_t]&:=& -\frac{i}{\hbar} [H_{\calD},\rho_{t}] \label{Disnd} ,
\eeqa
with
\begin{equation}
    H_{\mathcal{D}}:= i\sum_{m\neq n}\frac{\braketav{m}{\mathcal{D}[\rho_t]}{n}}{p_n-p_m}\ketbra{m}{n}
\end{equation}
being the ``bath Hamiltonian" that describes unitary part of the time-evolution induced by the bath~\cite{Funo-Speed-2019}.

Using Eqs.~(\ref{disd}) and (\ref{Disnd}), the non-adiabatic work~(\ref{nonadW}) reads 
\beqa
    |\dot{W}-\dot{W}_{\rm ad}| &=& \Bigl| \Tr\Bigl[\Linvad[\dot{H}_{t}] \Bigl( -\frac{i}{\hbar}[H_{t}+H_{\calD},\rho_{t}] + \calD_{t}[\rho_{t}]\Bigr) \Bigr] \Bigr| \\
    &\leq& \frac{1}{\hbar}\left|\Tr\left[ \left[\tilde{\mathcal{L}}_t^+[\dot{H}_t],H_t+H_\mathcal{D}\right]\rho_t\right]\right| + \left|\Tr[\tilde{\mathcal{L}}_t^+[\dot{H}_t]\mathcal{D}_\mathrm{d}[\rho_t]]\right|, \label{WnadboundA1}
\eeqa
where we use the triangle inequality and obtain the second line. 

The first term on the right-hand side of~(\ref{WnadboundA1}) can be further bounded as 
\beqa
\frac{1}{\hbar}\left|\Tr\left[ \left[\tilde{\mathcal{L}}_t^+[\dot{H}_t],H_t+H_\mathcal{D}\right]\rho_t\right]\right| &\leq& \frac{1}{\hbar} \sqrt{ \mathcal{F}_{\rho_{t}}(H_{t}+H_{\calD})V_{\rho_{t}}(\Linvad[\dot{H}_{t}])} \nonumber \\
&=& \sqrt{ 4g_{tt}^{\rm QF} (\Linvad[\dot{H}_{t}])} =: \mathcal{B}_{1}^{\rm nd}, \label{WnadboundA2}
\eeqa
which is shown in Appendix~\ref{appendix:firstbound}. On the other hand, the second term on the right-hand side of~(\ref{WnadboundA1}) can be further bounded as: 
\begin{equation}
\left|\Tr[\tilde{\mathcal{L}}_t^+[\dot{H}_t]\mathcal{D}_\mathrm{d}[\rho_t]]\right|\leq\sqrt{2A(\rho_{t})\dot{\sigma}} =: \mathcal{B}_{1}^{\rm d}, \label{WnadboundA4}
\end{equation}
which is shown in Appendix~\ref{appendix:secondbound}. By combining Eqs.~(\ref{WnadboundA1}), (\ref{WnadboundA2}), and (\ref{WnadboundA4}), we obtain
\beq
|\dot{W}-\dot{W}_{\rm ad}| \leq \sqrt{4g_{tt}^{\rm QF}V_{\rho_t}(\Linvad[\dot{H}_t])} + \sqrt{ 2A(\rho_{t})\dot{\sigma}},
\eeq
which proves the first main result~(\ref{mainresult}). 

%In the following, we bound from above each term that appears on the right-hand side of~(\ref{WnadboundA1}).

\subsection{\label{appendix:firstbound}Proof of~(\ref{WnadboundA2})}
%We now derive the following inequality and bound the first term on the right-hand side of~(\ref{WnadboundA1}): 
%\beqa
%\frac{1}{\hbar}\left|\Tr\left[ \left[\tilde{\mathcal{L}}_t^+[\dot{H}_t],H_t+H_\mathcal{D}\right]\rho_t\right]\right| &\leq& \frac{1}{\hbar} \sqrt{ \mathcal{F}_{\rho_{t}}(H_{t}+H_{\calD})V_{\rho_{t}}(\Linvad[\dot{H}_{t}])} \nonumber \\
%&=& \sqrt{ g_{tt}^{\rm QF} (\Linvad[\dot{H}_{t}])}. \label{WnadboundA2}
%\eeqa

We now derive inequality~(\ref{WnadboundA2}). The first line of~(\ref{WnadboundA2}) follows from the inequality~\cite{PhysRevX.12.011038, sekiguchi2024}
\beq
|\Tr( [X,Y]\rho_{t} ) | \leq \sqrt{ \mathcal{F}_{\rho_{t}}(X)V_{\rho_{t}}(Y)}, \label{coherencevariance}
\eeq
where the quantum Fisher information is defined as
\begin{equation}
    \mathcal{F}_{\rho}(X):=2\sum_{m\neq n}\frac{(p_n-p_m)^2}{p_n+p_m}\left|\braketav{n}{X}{m}\right|^2.
\end{equation}
To show inequality~(\ref{coherencevariance}), we use the spectral decomposition of the density operator as $\rho_{t}=\sum_{n}p_{n}|n\rangle\langle n|$ and obtain
\beqa
|\Tr( [X,Y]\rho_{t} ) | &=& \Bigl| \sum_{m\neq n}(p_{n}-p_{m})\langle n|X|m\rangle\langle m|Y|n\rangle \Bigr| \nonumber \\
&\leq & 2\sum_{n\neq m}\frac{(p_{n}-p_{m})^{2}}{p_{n}+p_{m}}|\langle n|X|m\rangle|^{2} \frac{1}{2}\sum_{n\neq m}(p_{n}+p_{m})|\langle m|Y|n\rangle|^{2} , \label{coherencevariance1}
\eeqa
where we use the Cauchy-Schwartz inequality and obtain the second line. We further note that 
\beqa
 \frac{1}{2}\sum_{n\neq m}(p_{n}+p_{m})|\langle m|Y|n\rangle|^{2} &=&  \Tr[Y^{2}\rho_{t}]-\sum_{n}p_{n}|\langle n|Y|n\rangle|^{2}  \nonumber \\
&\leq &  \Tr[Y^{2}\rho_{t}]-\Bigl(\sum_{n}p_{n}|\langle n|Y|n\rangle|\Bigr)^{2}  \nonumber \\
&=& V_{\rho_{t}}[Y]. \label{coherencevariance2}
\eeqa
By combining~(\ref{coherencevariance1}) and (\ref{coherencevariance2}) and using the definition of the quantum Fisher information, we obtain Eq.~(\ref{coherencevariance}). 

The second line of Eq.~(\ref{WnadboundA2}) follows from the relation
\beq
\langle n|\dot{\rho}_{t}|m\rangle = -\frac{i}{\hbar}(p_{m}-p_{n})\langle n|(H_{t}+H_{\calD})|m\rangle =   (p_{m}-p_{n})\langle n |\partial_{t}m\rangle   , \label{WnadboundA3}
\eeq
for $n\neq m$. Using this expression~(\ref{WnadboundA3}), we find that 
\beq
\frac{1}{\hbar^{2}}\mathcal{F}_{\rho_{t}}(H_{t}+H_{\calD}) = 4g_{tt}^{\rm QF} , \label{QFI=QFIM}
\eeq
completing the proof of~(\ref{WnadboundA2}). 

\subsection{\label{appendix:remarkQFIM}Remarks on the quantum Fisher information metric}
In this subsection, we give several remarks on the quantum Fisher information metric. Let us consider the Bures angle, which is the geodesic distance related to the quantum Fisher information metric, defined as~\cite{PhysRevX.6.021031, nielsen2010quantum}
\beq
\mathcal{L}(\rho,\sigma):=\arccos{\sqrt{F(\rho,\sigma)}},
\eeq
where
\beq
F(\rho,\sigma):=\Bigl( \Tr\Bigl[\sqrt{\sqrt{\rho}\sigma\sqrt{\rho}}\Bigr]\Bigr)^{2}
\eeq
is the Uhlmann fidelity. By evaluating the fidelity between $\rho_{t}$ and $\rho_{t}+d\rho_{t}$ and denoting $d\rho_{t}=\dot{\rho}_{t}dt$, we find up to the lowest order in $dt$ that~\cite{HUBNER1992239} 
\beq
\mathcal{L}(\rho_{t},\rho_{t}+d\rho_{t}) = \sqrt{ (g_{tt}^{\rm CL}+g_{tt}^{\rm QF})dt^{2}} , \label{metric}
\eeq
where
\beq
g_{tt}^{\rm CL}:= \frac{1}{4}\sum_{j} \frac{(\partial_{t}p_{j})^{2}}{p_{j}}
\eeq
is the classical Fisher information metric and $g_{tt}^{\rm QF}$ defined in Eq.~(\ref{QFIM}) describes a purely quantum contribution to Eq.~(\ref{metric}).

\subsection{\label{appendix:secondbound}Proof of~(\ref{WnadboundA4})}
To show~(\ref{WnadboundA4}), we start by introducing 
\begin{equation}
    a_n:=\langle n|\tilde{\mathcal{L}}_t^+[\dot{H}]|n\rangle ,
\end{equation}
and a classical transition rate-like quantity~\cite{Funo-Speed-2019}
\begin{equation}    W_{m,n}^{\omega}:=\gamma(\omega)\left|\braketav{m}{L_{\omega}}{n}\right|^2,
\end{equation}
and rewrite the left-hand side of~(\ref{WnadboundA4}) as 
\beq
\left|\Tr[\tilde{\mathcal{L}}_t^+[\dot{H}_t]\mathcal{D}_\mathrm{d}[\rho_t]]\right| = \Bigl| \sum_{m} a_{m} \sum_{n,\omega}(W_{m,n}^{\omega}p_n-W_{n,m}^{-\omega}p_m) \Bigr|. \label{WnadboundA5}
\eeq
We then use the following relation   
\beqa
\Bigl| \sum_{m} a_{m} \sum_{n,\omega}(W_{m,n}^{\omega}p_n-W_{n,m}^{-\omega}p_m) \Bigr|^{2} &\leq & \sum_{m,n,\omega}a_m^2(W_{m,n}^{\omega}p_n+W_{n,m}^{-\omega}p_m)\sum_{m,n,\omega}\frac{(W_{m,n}^{\omega}p_n-W_{n,m}^{-\omega}p_m)^2}{W_{m,n}^{\omega}p_n + W_{n,m}^{-\omega}p_m}\nonumber \\
&\leq& 2A(\rho_{t})\cdot \frac{1}{2}\sum_{m,n,\omega}(W_{m,n}^{\omega}p_n-W_{n,m}^{-\omega}p_m)\ln\frac{W_{m,n}^{\omega}p_n}{W_{n,m}^{-\omega}p_m} \nonumber \\
&=& 2A(\rho_{t}) \dot{\sigma}, \label{WnadboundA6}
\eeqa
where the first line is obtained by using the Cauchy-Schwarz inequality, and the second line is obtained by using the relation $2(a-b)^2/(a+b)\leq (a-b)\ln\frac{b}{a}$ for $a,b\geq 0$. In Eq.~(\ref{WnadboundA6}), we note that
\begin{equation}
A(\rho_{t})=\frac{1}{2}\sum_{m,n,\omega}a_m^2(W_{m,n}^{\omega}p_n+W_{n,m}^{-\omega}p_m) ,
\end{equation}
which follows from 
\beq
\sum_{n,\omega}W_{m,n}^{\omega}p_n= \sum_{\omega}\gamma(\omega)\langle m|L_{\omega}\rho_{t}L_{\omega}^{\dagger}|m\rangle
\eeq
and 
\beq
\sum_{n,\omega}W_{n,m}^{-\omega}p_m= \sum_{\omega} \gamma(\omega)\langle m|\rho_{t}L_{\omega}^{\dagger}L_{\omega}|m\rangle=\sum_{\omega} \gamma(\omega)\langle m|L_{\omega}^{\dagger}L_{\omega}\rho_{t}|m\rangle.
\eeq
We also use the following expression for the entropy production rate~\cite{Funo-Speed-2019}
\beq
\dot{\sigma} = \frac{1}{2}\sum_{m,n,\omega}(W_{m,n}^{\omega}p_n-W_{n,m}^{-\omega}p_m)\ln\frac{W_{m,n}^{\omega}p_n}{W_{n,m}^{-\omega}p_m} ,
\eeq
and derive the last line of~(\ref{WnadboundA6}). We finally obtain Eq.~(\ref{WnadboundA4}) by combining Eqs.~(\ref{WnadboundA5}) and (\ref{WnadboundA6}).

\section{\label{appendix:secondproof}Proof of the second main result}
In this section, we derive the second main result, Eqs.~(\ref{Wclbound}) and (\ref{Wqmbound}). 

Let us first derive Eq.~(\ref{Wqmbound}). Starting from the expression Eq.~(\ref{WqmHinv}), we have
\beq
|\dot{W}_{\rm qm}|=\frac{1}{\hbar}\left|\Tr\left[ \left[ \Hinvad(\dot{H}_{t}^{\rm qm}),H_{t}\right]\rho_{t} \right]\right| \leq \frac{1}{\hbar}\sqrt{ \mathcal{F}_{\rho_{t}}(H_{t})V_{\rho_{t}}(\Hinvad[\dot{H}_{t}^{\rm qm}])}, 
\eeq
where we used~(\ref{coherencevariance}) and obtain the last inequality.  
By further using Eq.~(\ref{Hinvad=Hcd}), we complete the proof of Eq.~(\ref{Wqmbound}). 

Next, we derive Eq.~(\ref{Wclbound}). We first note that the following relation holds because $\calD_{t}$ does not create coherence between energy eigenstates:
\beq
\calD_{t}[\rho_{t}^{\rm cl}]=\sum_{n}|\epsilon_{n}\rangle\langle \epsilon_{n}| \calD_{t}[\rho_{t}^{\rm cl}] |\epsilon_{n}\rangle\langle \epsilon_{n}| \label{Drhodiag} .
\eeq
%\beq
%\langle \epsilon_{n}|\Dinvad[\dot{H}_{t}^{\rm cl}]|\epsilon_{m}\rangle = 0, \ \text{ for } n\neq m,
%\eeq
%because $\calD_{t}$ do not create coherence between energy eigenstates. Therefore, the following relation holds:
%\beq
%\Dinvad[\dot{H}_{t}^{\rm cl}] = \sum_{n}|\epsilon_{n}\rangle\langle \epsilon_{n}| \Dinvad[\dot{H}_{t}^{\rm cl}] |\epsilon_{n}\rangle\langle \epsilon_{n}| \label{Dinvaddiag} .
%\eeq
We now start from the expression Eq.~(\ref{Wclbound}) and obtain
\beqa
|\dot{W}_{\rm cl}-\dot{W}_{\rm ad}|^{2} &=& |\Tr[ \Dinvad[\dot{H}_{t}^{\rm cl}]\calD_{t}[\rho_{t}^{\rm cl}]] |^{2} \nonumber \\
&=& \Bigl| \sum_{n} \langle \epsilon_{n}| \Dinvad[\dot{H}_{t}^{\rm cl}] |\epsilon_{n}\rangle \langle \epsilon_{n}|\calD_{t}[\rho_{t}^{\rm cl}]|\epsilon_{n}\rangle  \Bigr|^{2} \nonumber \\
&=& \Bigl| \sum_{n} \langle \epsilon_{n}| \Dinvad[\dot{H}_{t}^{\rm cl}] |\epsilon_{n}\rangle \sum_{\omega,m}( M_{nm}^{\omega}q_{m}-M_{mn}^{-\omega}q_{n})  \Bigr|^{2} \nonumber \\
&\leq & \Bigl(\sum_{n,m,\omega}\langle \epsilon_{n}| \Dinvad[\dot{H}_{t}^{\rm cl}] |\epsilon_{n}\rangle^{2}(M_{nm}^{\omega}q_{m}+M_{mn}^{-\omega}q_{n}\Bigr)\sum_{n,m,\omega}\frac{(M_{nm}^{\omega}q_{m}-M_{mn}^{-\omega}q_{n})^{2}}{M_{nm}^{\omega}q_{m}+M_{mn}^{-\omega}q_{n}} \nonumber \\
&\leq & \sum_{n,\omega}\langle \epsilon_{n}| \Dinvad[\dot{H}_{t}^{\rm cl}] |\epsilon_{n}\rangle^{2}\gamma(\omega)\langle \epsilon_{n}| L_{\omega}\rho_{t}^{\rm cl}L_{\omega}^{\dagger}+\rho_{t}^{\rm cl}L_{\omega}^{\dagger}L_{\omega}|\epsilon_{n}\rangle \nonumber \\
& & \times \frac{1}{2}\sum_{n,m,\omega}(M_{nm}^{\omega}q_{m}-M_{mn}^{-\omega}q_{n})\ln\frac{M_{nm}^{\omega}q_{m}}{M_{mn}^{-\omega}q_{n}} \nonumber \\
&=& 2A(\rho_{t}^{\rm cl}) \dot{\sigma}_{\rm cl}, 
\label{Wclbound_derivation}
\eeqa
where we have introduced 
\beq
M_{nm}^{\omega}:=\gamma(\omega)|\langle \epsilon_{n}|L_{\omega}|\epsilon_{m}\rangle|^{2}
\eeq
and 
\beq
q_{n}:=\langle\epsilon_{n}|\rho_{t}^{\rm cl}|\epsilon_{n}\rangle.
\eeq
Note that Eq.~(\ref{Wclbound_derivation}) is similar to Eq.~(\ref{WnadboundA6}). However, the entropy production rate is now given by
\beqa
& &\frac{1}{2}\sum_{n,m,\omega}(M_{nm}^{\omega}q_{m}-M_{mn}^{-\omega}q_{n})\ln\frac{M_{nm}^{\omega}q_{m}}{M_{mn}^{-\omega}q_{n}} \nonumber \\
&=& \sum_{n,m,\omega}M_{nm}^{\omega}q_{m}\ln\frac{M_{nm}^{\omega}q_{m}}{M_{mn}^{-\omega}q_{n}} \nonumber \\
&=& \sum_{n,m,\omega}M_{nm}^{\omega}q_{m}\ln\frac{\gamma(\omega)q_{m}}{\gamma(-\omega)q_{n}} \nonumber \\
&=& \beta\sum_{\omega}\omega\gamma(\omega) \Tr[L_{\omega}^{\dagger}L_{\omega}\rho_{t}^{\rm cl}] + \sum_{\omega}\gamma(\omega)\Tr[ L_{\omega}^{\dagger}L_{\omega}\rho_{t}^{\rm cl}\ln\rho_{t}^{\rm cl}-\ln\rho_{t}^{\rm cl}L_{\omega}\rho_{t}^{\rm cl}L_{\omega}^{\dagger}] \nonumber \\
&=& \beta \dot{Q} - \Tr[(\ln\rho_{t}^{\rm cl})\calL_{t}[\rho_{t}^{\rm cl}]] \nonumber \\
&=& \dot{\sigma}_{\rm cl}. 
\eeqa
%and obtain the final equality of~(\ref{Wclbound_derivation}) is obtained. 
This completes the proof of Eq.~(\ref{Wclbound}). 

In the following, we show $\dot{\sigma}_{\rm cl}\leq \dot{\sigma}$ based on Ref.~\cite{Tajima-Superconducting-like-2021}. We first introduce a completely-positive and trace-preserving (CPTP) map 
\beq
\mathcal{E}_{t}[\chi] :=\sum_{n}|\epsilon_{n}(t)\rangle\langle \epsilon_{n}(t)| \chi |\epsilon_{n}(t)\rangle\langle \epsilon_{n}(t)|,
\eeq
which projects the density operator to diagonal states, i.e., $\rho_{t}^{\rm cl}=\mathcal{E}_{t}[\rho_{t}]$. We next introduce a CPTP map that describes the infinitesimal time-evolution $t\rightarrow t+\Delta t$ as 
\beq
\Lambda_{t,t+\Delta t} := (1+ \calL_{t}\Delta t),
\eeq
and denote $\rho_{t+\Delta t}=\Lambda_{t,t+\Delta t} [\rho_{t}]$ and $\rho_{t+\Delta t}^{\rm cl}=\Lambda_{t,t+\Delta t}[ \rho_{t}^{\rm cl}]$.  It should be noted that $\Lambda_{t,t+\Delta t}[\mathcal{E}_{t}[\chi]]=\mathcal{E}_{t}[\Lambda_{t,t+\Delta t}[\chi]]$, and therefore $\rho_{t+\Delta t}^{\rm cl}=\mathcal{E}_{t}[\rho_{t+\Delta t}]$.  
We now express the difference between $\dot{\sigma}$ and $\dot{\sigma}_{\rm cl}$ as 
\beqa
\dot{\sigma}-\dot{\sigma}_{\rm cl}&=& \lim_{\Delta t\rightarrow 0}\frac{1}{\Delta t} \Bigl( S(\rho_{t+\Delta t}) - S(\rho_{t}) - S(\rho^{\rm cl}_{t+\Delta t}) + S(\rho^{\rm cl}_{t})\Bigr) \nonumber \\
&=& \lim_{\Delta t\rightarrow 0}\frac{1}{\Delta t} \Bigl(\Tr\Bigl[\rho_{t}\Bigl(\ln\rho_{t} - \ln \mathcal{E}_{t}[\rho_{t}]\Bigr) \Bigr] -\Tr\Bigl[\rho_{t+\Delta t}\Bigl(\ln \rho_{t+\Delta t} -\ln \mathcal{E}_{t}[\rho_{t+\Delta t}] \Bigr) \Bigr] \Bigr) \nonumber \\
&=& \lim_{\Delta t\rightarrow 0}\frac{1}{\Delta t} \Bigl( D(\rho_{t} || \mathcal{E}_{t}[\rho_{t}]) - D(\Lambda_{t,t+\Delta t}[\rho_{t}] || \Lambda_{t,t+\Delta t}[\mathcal{E}_{t}[\rho_{t}]]) \Bigr) \geq 0, \label{sigma-sigmacl}
\eeqa
where we used $\Tr[\rho_{t}\ln \mathcal{E}_{t}[\rho_{t}]]=\Tr[\rho_{t}^{\rm cl}\ln\rho_{t}^{\rm cl}]$ and obtained the second line, $D(\rho||\sigma):=\Tr[\rho(\ln\rho -\ln \sigma)]$ is the quantum relative entropy, and the last inequality follows from the monotonicity of the relative entropy~\cite{nielsen2010quantum}. 

\section{\label{appendix:example}Details of the example}
In this section, we show additional details of the two-level system example discussed in Sec.~\ref{sec:example}.
%In the following, we use the notation $p_{\pm}=\langle \epsilon_{\pm}|\rho|\epsilon_{\pm}\rangle$ and $\rho_{+-}=\langle \epsilon_{+}|\rho|\epsilon_{-}\rangle$, %The energy eigenstates are given by $|\epsilon_{+}\rangle=\cos\theta_{t}|e\rangle + \sin\theta_{t}|g\rangle$ and $|\epsilon_{-}\rangle=\sin\theta_{t}|e\rangle-\cos\theta_{t}|g\rangle$, where $|e\rangle$ and $|g\rangle$ are the eigenstates of $\sigma_{z}$, i.e., $\sigma_{z}|e\rangle=|e\rangle$ and $\sigma_{z}|g\rangle=-|g\rangle$. We also introduce $\theta_{t}=(1/2)\text{cot}^{-1}(q_{t}/\Delta_{t})$. 
and give explicit expressions of the non-adiabatic work and the bounds for two-level systems.
\begin{itemize}

\item The classical and quantum part of the non-adiabatic work reads
\beqa
\dot{W}_{\rm cl}-\dot{W}_{\rm ad} &=& \frac{\dot{\omega}_{t}}{2}(p_{+}-p_{-}) - \frac{\dot{\omega}_{t}}{2}\frac{\gamma_{\uparrow}-\gamma_{\downarrow}}{\gamma_{\uparrow}+\gamma_{\downarrow}}, \\ \dot{W}_{\rm qm} &=& -2\omega_{t}\dot{\theta}_{t}\Re[\rho_{+-}].
\eeqa

%\item The counter-diabatic Hamiltonian is given by
%    \beq
%    H_{\rm cd}= i\dot{\theta}_{t}(|\epsilon_{+}\rangle\langle \epsilon_{-}|-|\epsilon_{-}\rangle\langle\epsilon_{+}|)=\dot{\theta}_{t}\sigma_{y},
%    \eeq
%and its variance reads
%   \beq
%    V_{\rho_{t}}(H_{\rm cd})=\dot{\theta}^{2}_{t}\left(1-4(\Im[\rho_{+-}])^{2}\right).
%    \eeq

\item The diagonal entropy production reads
    \beq
    \dot{\sigma}_{\rm cl}= (\gamma_{\downarrow}p_{+}-\gamma_{\uparrow}p_{-}) \Bigl( \beta\omega +\ln\frac{p_{+}}{p_{-}} \Bigr).  
    \eeq
%\item The activity-like quantity is given by
%    \beq
%    A(\rho_{t}^{\rm cl}) = \frac{2\dot{\omega}_{t}^{2}A_{\rm act}}{(\gamd+\gamu)^{2}}, \label{Adiag_ex}
%    \eeq
%    where
%    \beq
%A_{\rm act}=\gamma_{\uparrow}p_{-}+\gamma_{\downarrow}p_{+}
%\eeq
%is the activity. 
    
\end{itemize}

When deriving Eq.~(\ref{Adiag_ex}), we use the following explicit form of the the Drazin inverse~\cite{PhysRevE.104.034117} and its adjoint:
\begin{equation}
\begin{split}
    \mathcal{L}^+[\rho]=&
     -\dfrac{\gamd}{(\gamd+\gamu)^2}\ketbra{\epsilon_+}{\epsilon_+}\rho\ketbra{\epsilon_+}{\epsilon_+}
    +\dfrac{\gamd}{(\gamd+\gamu)^2}\ketbra{\epsilon_-}{\epsilon_+}\rho\ketbra{\epsilon_+}{\epsilon_-}\\
    &+\dfrac{\gamu}{(\gamd+\gamu)^2}\ketbra{\epsilon_+}{\epsilon_-}\rho\ketbra{\epsilon_-}{\epsilon_+}
    -\dfrac{\gamu}{(\gamd+\gamu)^2}\ketbra{\epsilon_-}{\epsilon_-}\rho\ketbra{\epsilon_-}{\epsilon_-}\\
    &-\dfrac{1}{(\gamd+\gamu)/2+i\omega}\ketbra{\epsilon_+}{\epsilon_+}\rho\ketbra{\epsilon_-}{\epsilon_-}
    -\dfrac{1}{(\gamd+\gamu)/2-i\omega}\ketbra{\epsilon_-}{\epsilon_-}\rho\ketbra{\epsilon_+}{\epsilon_+},
\end{split}
\end{equation}
and
\begin{equation}
\begin{split}
    \tilde{\mathcal{L}}^+[A]=&
    -\dfrac{\gamd}{(\gamd+\gamu)^2}\ketbra{\epsilon_+}{\epsilon_+}A\ketbra{\epsilon_+}{\epsilon_+}
    +\dfrac{\gamd}{(\gamd+\gamu)^2}\ketbra{\epsilon_+}{\epsilon_-}A\ketbra{\epsilon_-}{\epsilon_+}\\
    &+\dfrac{\gamu}{(\gamd+\gamu)^2}\ketbra{\epsilon_-}{\epsilon_+}A\ketbra{\epsilon_+}{\epsilon_-}
    -\dfrac{\gamu}{(\gamd+\gamu)^2}\ketbra{\epsilon_-}{\epsilon_-}A\ketbra{\epsilon_-}{\epsilon_-}\\
    &-\dfrac{1}{(\gamd+\gamu)/2+i\omega}\ketbra{\epsilon_-}{\epsilon_-}A\ketbra{\epsilon_+}{\epsilon_+}
    -\dfrac{1}{(\gamd+\gamu)/2-i\omega}\ketbra{\epsilon_+}{\epsilon_+}A\ketbra{\epsilon_-}{\epsilon_-}. \label{Ladinv_twolevel}
\end{split}
\end{equation}

From the expression~(\ref{Ladinv_twolevel}), we find that
\beq
\tilde{\mathcal{L}}^+[\dot{H}_{t}]=-\frac{\dot{\omega}_{t}}{\gamd+\gamu}\Bigl( |\epsilon_{+}\rangle\langle \epsilon_{+}|-|\epsilon_{-}\rangle \langle \epsilon_{-}|\Bigr)  +\dfrac{\omega_{t}\dot{\theta}_{t}}{(\gamd+\gamu)/2+i\omega_{t}}|\epsilon_{-}\rangle\langle \epsilon_{+}| +\dfrac{\omega_{t}\dot{\theta}_{t}}{(\gamd+\gamu)/2-i\omega_{t}}|\epsilon_{+}\rangle\langle \epsilon_{-}|. \label{LinvadH_ex}
\eeq

\end{document}